\documentstyle[preprint,prc,aps]{revtex}
\newcommand{\LA}[1]{\mbox{\LARGE $#1$}}
\begin{document}
\preprint{
\rightline{\vbox{\hbox{\rightline{McGill/94-56}}
\hbox{\rightline{MSUCL-961}}}}
         }

\title{Another look at ``soft'' lepton pair production \\
in nucleon-nucleon bremsstrahlung}
\author{Jianming Zhang\thanks{Electronic address:
jzhang@hep.physics.mcgill.ca}, Rahma Tabti\thanks{Electronic address:
tabti@hep.physics.mcgill.ca}, and
Charles Gale\thanks{Electronic address: gale@hendrix.physics.mcgill.ca}}
\address{
Physics Department, McGill University, Montr\'eal,
Qu\'ebec, Canada H3A--2T8}
\author{Kevin Haglin\thanks{Electronic address: haglin@theo03.nscl.msu.edu}}
\address{
National Superconducting Cyclotron Laboratory, Michigan State University\\
East Lansing, MI 48824-1321 USA}
\maketitle
\begin{abstract}
We compare different formalisms for the calculation of lepton pair
emission in hadron-hadron collisions and discuss the consequences of the
approximations inherent to each of them.
Using a Lorentz-covariant and gauge-invariant formalism proposed by
Lichard, we calculate lepton pair emission via bremsstrahlung in
proton-proton and neutron-proton reactions at energies between 1 and
5 GeV to leading order in the virtual photon four-momentum.
These new results
suggest that some previous bremsstrahlung calculations based on varieties
of the
soft-photon approximation  might have somewhat overestimated
the  DLS low-mass dilepton cross sections.
We find that the respective
intensities of dilepton production through $p p$ and $n p$ bremsstrahlung
are energy-dependent and become
comparable at $E_{\rm kin} \agt $ 2 GeV. We also calculate the polar
angle anisotropy of the lepton spectrum.
\end{abstract}
\section{Introduction}
Dileptons and photons are probably the best carriers of information from the
hot and  compressed nuclear matter produced in the early stages of heavy-ion
collisions~\cite{gsi94}. In principle, those electromagnetically
interacting particles can leave the hadronic environment from which they are
created without significant disturbances, offering a relatively clean probe
of the nuclear collision dynamics.  In heavy ion collisions at incident
energies
of 1--5  $A$ GeV, where the Lawrence Berkeley Laboratory Dilepton Spectrometer
(DLS) has already taken measurements~\cite{dls}, the most
important sources of $e^+e^-$ pairs seem to be Dalitz and radiative
decays---mainly from $\eta$ mesons and $\Delta$'s,
pion-pion annihilations, and nucleon-nucleon
bremsstrahlung~\cite{xia90,wg90,haglin94}.  Measurements of lepton
pair production in
single nucleon-nucleon reactions have also recently been performed in
the GeV energy regime.
In those, the
measured pd/pp dielectron yield ratios display a clear beam energy as
well as invariant mass dependence~\cite{wkw93}. This suggests that
the dominant mechanism for dilepton production may be changing as  the
beam energy per nucleon  increases from 1 to 5 GeV. In particular, one
should pay
attention to the opening of inelastic nucleon-nucleon channels. The latter
have been
shown to play an important role for dielectron production in
nucleon-nucleon collisions
at 4.9 GeV \cite{haglin94}.     To eventually understand
quantitatively and completely the relative role
of all these contributions and their excitation function
in the complex environment of
nucleus-nucleus collisions, it is vital to first calculate the
lepton pair production cross sections for individual processes  as
accurately as
possible. Here we shall concentrate on the
bremsstrahlung generation of lepton pairs in the case of nucleon-nucleon
reactions.

Several different calculations for electron-positron pair
emission through nucleon-nucleon
bremsstrahlung  have been
performed for reactions at, and slightly above, 1
GeV.  Some of the more
sophisticated approaches used relativistic one-boson exchange (OBE)
Lagrangians, with the
coupling to the electromagnetic field done by minimal substitution
\cite{hkg89,sbcm89,kev91}. These approaches are thus entirely
Lorentz-covariant
and are also gauge-invariant in the electromagnetic sector. Note that the OBE
dilepton calculations can be made gauge-invariant even when form factors at
the strong interactions vertices are used \cite{hkg89}. Such approaches
have all the
transformation properties
that are required of a complete theory but they are not completely
satisfactory in two
respects. First, they are very cumbersome. If several meson fields are
involved, the
number of Feynman graphs to be evaluated proliferates rapidly and the
difficulty of the
calculations increases accordingly. The coupling constants in the OBE
model are fitted
such that the total nucleon-nucleon cross sections are reproduced
as closely as
possible. This  exercise thus has to be repeated for each
colliding system. Second, even if two different OBE calculations with
two slightly
different set of ingredients (meson fields, form factors and coupling
constants) can do
a good job of generating total nucleon-nucleon cross sections that are
in agreement
with experimental measurements, generally they will have different
predictions for
the differential cross sections. As we shall see below, there is a way
of writing the
low invariant mass dilepton production cross section in nucleon-nucleon
collisions such that it clearly depends on the differential elastic
cross section of
the colliding partners. This fact thus imposes very stringent
requirements on the OBE
models as far as their ability to predict lepton pair production yields
is concerned.
This point was
recently made in the literature \cite{law93,haglin94}. Because of the above
considerations, the calculations pertaining to the bremsstrahlung generation of
low invariant mass lepton pairs in nucleon-nucleon collisions have used
a ``soft photon approximation'' \cite{bd64} of some kind or another.
Almost all of the recent
calculations of
dielectron
production in nucleon-nucleon (and nucleus-nucleus) collisions
that have used the soft
photon approximation
have used as their starting point a formula suggested by R\"uckl \cite{rr76}:
\begin{equation}
E_+ E_- {{d^6 \sigma^{e^+ e^-}}\over {d^3 p_+ d^3 p_-}}\  = \ {{\alpha}\over {2
\pi^2}}
{{1}\over{M^2}} \left( \omega {{d^3 \sigma^\gamma}\over{d^3 q}}
\right)_{\vec{q} = \vec{p_+} + \vec{p_-}}\ .
\end{equation}
This equation links the cross section for production of dileptons via
virtual photon
bremsstrahlung to the bremsstrahlung cross section for real photons. In
the above,
$\vec{p_\pm}$ is the three-momentum of the electron or positron, $E_\pm$
is the energy,
$M^2 \ =\ (p_+ + p_-)^2$ is the dilepton invariant mass squared,
$\vec{q}$ is the
photon momentum and $\omega$ is its energy. The fine structure constant
appears as
$\alpha$.

The derivation of soft photon formul{\ae} in the
context of bremsstrahlung emission of lepton pairs has recently been
re-analyzed \cite{pl94}. It was shown that R\"uckl's formula was not properly
Lorentz-covariant  and did not contain the relationship between dilepton
cross section
and virtual photon cross section that is required by gauge
invariance \cite{bpp91}. In
the context of the new analysis  of dilepton production via virtual photon
bremsstrahlung, a new set of formul{\ae} were derived for the leading
order and
next-to-leading order contributions to lepton pair emission \cite{pl94}.

It had been
known for some
time that the R\"uckl formula was not entirely
correct~\cite{gale87,hkg89,sbcm89,kev91}. Nevertheless, in  comparison
with OBE
calculations, it was
deemed quantitatively adequate~\cite{wg90,kev91}.  It is not the point
of this
paper to restate
this known fact. We rather wish to investigate  the empirical predictions
of the new
formalism in the
practical framework of quasi-elastic nucleon-nucleon collisions
at energies around and above
1 GeV, to establish whether
the new leading term approximation deviates significantly from previous
works. We also plan to investigate
quantitatively and systematically the effects of different layers of
approximation found in several calculations.
Our paper is
organized as
follows: in section \ref{sec2} we introduce the general formalism,
in section \ref{sec3} we compare different approaches to the generation
of soft virtual photons and discuss the differences. In section
\ref{phasesp} we explore the consequences of a better treatment of the
many-body phase space. In section \ref{angular}, we discuss how some of
the approximations seen in sections \ref{sec2} and \ref{sec3} can be put
into perspective by an angular anisotropy analysis of the lepton
spectrum. We then summarize.

\section{General formalism and hadronic electromagnetic currents}
\label{sec2}

Consider the reaction
\begin{equation}
a + b \rightarrow c + d + e^{+}e^{-}
\label{eq:eq1}
\end{equation}
where $a$, $b$, $c$, and $d$ represent nucleons. The generic Feynman
diagram for
the leading-term contributions to the emission of a soft virtual
photon are shown
in Fig. \ref{fig1}. This is the process on
which we concentrate in this work. See the appendix for a mathematical
description of
what is a leading term and what isn't.
The circle schematically represents
the strong interaction. In the limit of soft photons, real or virtual,
the radiation
from the strong interaction blob is a sub-leading contribution.
When $a$, $b$, $c$ and $d$  are all protons, there are
eight Feynman graphs to be added coherently for the calculation of lepton pair
emission.  Note in passing that there have been arguments that
bremsstrahlung from $p n$  reactions should be significantly more important
than that from $p p$.   These were based on the fact that, nonrelativistically,
the first non-vanishing multipole contribution for $p p$ appears at the
quadrupole
level, whereas for $n p$ it is at the dipole stage.
It has however been recently shown that such arguments do not hold for
relativistic
collisions ~\cite{haglin94}. In fact, the lepton pair yields from these two
processes are comparable at 4.9 GeV~\cite{haglin94}. There will be more on
this later.  Naturally, the
$p p$ bremsstrahlung contribution is crucial in the interpretation of
the ratio  of
dileptons produced in $p d$ and $p p$ reactions.
We will discuss $p p$ contributions in this work, also.

We evaluate and sum the amplitudes of the relevant Feynman diagrams of
the type in Fig.
\ref{fig1}. Anticipating the soft photon limit, we assume that the hadronic
part of the total  matrix element is unaffected by the fact that one of
its legs is
slightly off-shell. We write this on-shell matrix element for elastic
nucleon-nucleon
scattering as ${\cal M}_0$. We omit the momentum labelling of the initial
and final
state, for simplicity.
Squaring the net matrix element, ${\cal M}$, and summing over final
state spins and averaging over initial states, we obtain to leading
order (see appendix)

\begin{equation}
\frac {1}{4} \sum_{s_a s_b s_c s_d s_+ s_-} \left |
{\cal M} \right |^2 \ =\ 4 \pi \alpha \overline{\left |
{\cal M}_0 \right
|^2} J^\mu J^\nu L_{\mu \nu}\ .
\label{eq:M2}
\end{equation}
In the above,
\begin{equation}
J^{\mu} = -Q_{a} {(2p_{a}-q)^{\mu} \over 2p_{a}\cdot q -M^{2}} -
Q_{b} {(2p_{b}-q)^{\mu} \over 2p_{b}\cdot q -M^{2}} +
Q_{c} {(2p_{c}+q)^{\mu} \over 2p_{c}\cdot q +M^{2}}
+ Q_{d} {(2p_{d}+q)^{\mu} \over 2p_{d}\cdot q +M^{2}}
\label{eq:jmu1}
\end{equation}
is the hadron electromagnetic current, and the lepton tensor is
\begin{equation}
L^{\mu \nu}\ =\ {{8 \pi \alpha}\over{M^4}} \left ( 2 \left ( {p_+}^\mu
{p_-}^\nu +
{p_-}^\mu {p_+}^\nu \right) - M^2 g^{\mu \nu} \right ) \ .
\label{eq:tensor}
\end{equation}
The $Q$'s and $p$'s  represent charges and four-momenta
for the particles, and $q^\mu = (p_+ + p_- )^\mu$.
Note that $J^{\mu}q_{\mu}=0$,
as a consequence of gauge invariance.   $\overline{| {\cal M}_0 |^2}$ is
the on-shell hadronic elastic scattering matrix element, squared, summed
over final spins and averaged over initial spins.

After performing the appropriate contractions, we may write
the differential cross
section for  $e^{+}e^{-}$ pair production with invariant
mass $M$ and energy $q_{0}$ as
\begin{equation}
E_{+}E_{-}\frac{d^{6}\sigma^{e^{+}e^{-}}_{ab\rightarrow
cd}}{d^{3}p_{+}d^{3}p_{-}} =
\frac{\alpha^{2}}{16\pi^5} \frac{1}{M^{2}}
\frac{R_{2}(s_{2},m^{2}_{a},m^{2}_{b})}{R_{2}(s,m^{2}_{a},m^{2}_{b})}
\int [-J^{2}-\frac{1}{M^{2}}(l \cdot J)^{2}]
\frac{d\sigma_{ab\rightarrow cd}}{dt}
d\phi^*_c dt \ ,
\label{eq:diff}
\end{equation}
where $\phi^*_c$ is the azimuthal angle of one of the outgoing
hadrons
in the center-of-mass frame and
the four-vector $l=p_+-p_-$ is the difference of positron and electron
four-momenta and $t$ is the four-momentum transfer.
Equation (\ref{eq:diff}) follows from Eq. (2.15) of \cite{pl94} after
specifying the nonradiative cross section as
\begin{equation}
d\sigma_0=
\frac{d\sigma_{ab\rightarrow cd}}{d\Omega_c}d\Omega_c
=\frac{1}{2\pi}\frac{d\sigma_{ab\rightarrow cd}}{dt}
d\phi^*_c dt \ ,
\end{equation}
and including the phase-space correction introduced in \cite{gale89}.

In the evaluation of the original Feynman diagrams, we have neglected the
four-momentum
$q$ of the virtual photon in the phase-space $\delta$ function in order
to reproduce the kinematics associated with the
{\em on-shell} elastic differential nucleon-nucleon cross section,
$d \sigma / d t$ \cite{bd64}. Because of this
approximation, we
include \cite{gale89} the ratio of two-body phase space~\cite{eb73}
$R_{2}(s_{2},m^{2}_{a},m^{2}_{b})/R_{2}(s,m^{2}_{a},m^{2}_{b})$
evaluated at $s_{2}$ and $s$, where $s$ is the invariant energy
squared available
for all the final-state particles and $s_{2}=s+M^{2}-2{q_{0}}^{*} \sqrt{s}$.
This approximate correction prevents the lepton pair from violating the
overall energy-momentum conservation laws and thus has a significant
effect on dilepton distributions. The ratio
constructed from Eq. (\ref{eq:diff}) with and without the correction factor
correctly drops monotonically to zero in the
limit of maximum invariant mass.  Handling the phase-space properly is
quite important. However, setting $q = 0$ in the four-dimensional delta
function also forces the expression for the current to be evaluated
on-shell, {\em
i.e.} $p_c + p_d = p_a + p_b$. Clearly, this will have some effect on
the value of the current of Eq. (\ref{eq:jmu1}). For the
purpose of clarity, we will investigate separately the effects of  a
complete treatment of phase space in section \ref{phasesp}.

An exercise in relativistic kinematics gets one from Eq. (\ref{eq:diff}) to the
Lorentz-covariant differential cross section in global dilepton variables
\begin{equation}
q_{0}\frac{d^{4}\sigma^{e^{+}e^{-}}_{ab \rightarrow cd}}{dM^{2}d^{3}q} =
\frac{\alpha^{2}}{24\pi^4} \frac{1}{M^{2}}(1+\frac{2\mu^{2}}{M^{2}})
\sqrt{1- \frac{4\mu^{2}}{M^{2}}}
\frac{R_{2}(s_{2},m^{2}_{a},m^{2}_{b})}{R_{2}(s,m^{2}_{a},m^{2}_{b})}
\int (-J^{2})\frac{d\sigma_{ab\rightarrow cd}}{dt}
d\phi^*_c dt
\label{eq:d6}
\end{equation}
where $\mu$ is the electron mass. Our usage of R\"uckl's approach for
dilepton production consists of
neglecting the term
$(p_{+}^{\mu}p_{-}^{\nu}+p_{+}^{\nu}p_{-}^{\mu})$
of the leptonic tensor \cite{sbcm89}, Eq. (\ref{eq:tensor}), and
analytically continuing the real photon energy to $q_0 = \sqrt{
\vec{q}^2 + M^2 }$. This procedure is however not Lorentz-covariant
\cite{pl94}. One then obtains Eq. (\ref{eq:d6})
multiplied by an additional
factor of  3/2 ( 1+ 2$\mu^{2}$/$M^{2})^{-1}$. From here on, we shall discuss
electron-positron pair production exclusively and we systematically will
use $\mu$~=~0.

To complete our analysis of previous efforts we now turn our attention to the
electromagnetic current.  For
emission of real photons in reaction (\ref{eq:eq1}), the current is
\begin{equation}
J^{\mu} = -Q_{a} {p_{a}^{\mu} \over p_{a}\cdot q} -
Q_{b} {p_{b}^{\mu} \over p_{b}\cdot q} +
Q_{c} {p_{c}^{\mu} \over p_{c}\cdot q}
+ Q_{d} {p_{d}^{\mu} \over p_{d}\cdot q}.
\label{eq:jmu}
\end{equation}
This approximation was also used by
several authors, in the case of soft {\em virtual} photons
\cite{gale87,xia90,xiong90,wg90,haglin94}. It is worth mentioning here
that for the case of virtual photons, the ``soft'' limit is not
unambiguously
defined: for virtual photons, what is meant by a ``soft photon limit''?
Here, we use this term in connection with the condition $M \rightarrow 0$.
However, bear in mind that the lepton pair energy and lepton pair
three-momentum
can individually still be quite large. Thus $M \rightarrow 0$ is in fact
not a sufficient condition to neglect both $q_0$ and $\vec{q}$ in the
delta functions, even if this omission can {\em partially} be corrected
({\em e.g.} Eq. (\ref{eq:diff})).   The opposite limit
where $M$ is large implies  that the lepton pair energy is large
and most certainly can't be ignored in the phase space delta
function.  By the same token, off-shell effects on the strong
interaction amplitude can be small in the $M \rightarrow 0$ limit, but
won't be, in the large $M$ region. In this work we do not include a
discussion of off-shell effects. We should however keep this point
in mind when discussing the regions of validity of our
calculations.

There are thus several possible ingredients that can affect the results
of lepton pair emission calculations of the ``soft photon type''.
One resides
in the use of
the R\"uckl formula, a widely-used expression which nevertheless is not
Lorentz-covariant. This first point is easily settled: our
interpretation of R\"uckl's
formula for electron-positron pair production instead of Eq.
(\ref{eq:d6}) simply introduces an overall factor of 3/2. Another
possible source of disagreement between calculations has to do with
the use of
the current of Eq.
(\ref{eq:jmu}) as the electromagnetic current for hadrons when
{\em massive} lepton pairs are
emitted. For lepton pairs with very small invariant masses  the
current represented by Eq.
(\ref{eq:jmu1}) of course
reduces to that of Eq. (\ref{eq:jmu}). Also, the exact treatment of
dynamics will play a role. For the time being, we set $q = 0$ in the
four-dimensional Dirac delta function and simply use the phase space
correction factor associated with Eq. (\ref{eq:diff}). In the next
section, we investigate the effects of using different hadronic
electromagnetic currents.

\section{Comparing currents}
\label{sec3}

Squaring the current (\ref{eq:jmu1}), as is required by Eq.
(\ref{eq:d6}), we obtain an expression that depends on the spatial
orientation of the
virtual photon. Taking an angular average, and  for an equal-mass
reaction of
the type specified in Eq. (\ref{eq:eq1})
with $m_{a}=m_{b}=m_{c}=m_{d}=m$, we arrive at
\begin{eqnarray}
-J^{2} &=& \left. {1\over q_{0}^{2}} \right\lbrace
-[\lambda_{1}^{2}(Q_{a}^{2}+Q_{b}^{2})
\frac{4m^{2}-M^{2}}{s(1- \beta^{'2}_{a})}
+\lambda_{2}^{2}(Q_{c}^{2}+Q_{d}^{2})
\frac{4m^{2}-M^{2}}{s(1- \beta^{'2}_{c})}] \nonumber\\
&\ &-
2\lambda_{1}^{2}Q_{a}Q_{b}(2-\frac{M^{2}}{s}-\frac{4m^{2}}{s})F(\vec
\beta^{\,'}_{a},\vec \beta^{\,'}_{b}) \nonumber\\
&\ &-
2\lambda_{2}^{2}Q_{c}Q_{d}(2-\frac{M^{2}}{s}-\frac{4m^{2}}{s})F(\vec
\beta^{\,'}_{c},\vec \beta^{\,'}_{d}) \nonumber\\
&\ &+
2\lambda_{1}\lambda_{2}(Q_{a}Q_{c}+Q_{b}Q_{d})\frac{4m^{2}+M^{2}-2t}{s}F(\vec
\beta^{\,'}_{a},\vec \beta^{\,'}_{c}) \nonumber\\
&\ & \left. +
2\lambda_{1}\lambda_{2}(Q_{a}Q_{d}+Q_{b}Q_{c})
\frac{2 s + M^2 - 4m^{2} + 2 t}{s}F(\vec
\beta^{\,'}_{a},\vec \beta^{\,'}_{d}) \right\rbrace,
\label{eq:correctedotj}
\end{eqnarray}
where the function
\begin{equation}
F(\vec x,\vec y\,)=\frac{1}{2\sqrt{R}}\ln \left| \frac{[\vec x \cdot \vec y
-x^{2}-\sqrt{R}][\vec x \cdot \vec y -y^{2}-\sqrt{R}]}{[\vec x \cdot \vec y
-x^{2}+\sqrt{R}][\vec x \cdot \vec y -y^{2}+\sqrt{R}]} \right|,
\end{equation}
and where the radicand is
\begin{equation}
 R=(1-\vec x \cdot \vec y\,)^{2}-(1-x^{2})(1-y^{2}).
\end{equation}
To make the formula more compact, we have introduced the variables
\begin{eqnarray}
&\ & \lambda_{1}=\frac{1}{1-M^{2}/\sqrt{s}q_{0}},  \; \; \;
\lambda_{2}=\frac{1}{1+M^{2}/\sqrt{s}q_{0}},  \; \; \;
\gamma=1-\frac{M^{2}}{q_{0}
^{2}}.
\label{eq:def}
\end{eqnarray}
The velocities $\beta$ are related to the above definitions and
to invariants through
\begin{eqnarray}
&\ & \beta^{'2}_{a}=\beta^{'2}_{b}=\lambda_{1}^{2}\gamma(1-\frac{4m^{2}}{s}),
     \; \; \;
\beta^{'2}_{c}=\beta^{'2}_{d}=\lambda_{2}^{2}\gamma(1-\frac{4m^{2}}{s}),
\nonumber\\
&\ & \vec \beta^{\,'}_{a} \cdot \vec
\beta^{\,'}_{b}=-\lambda_{1}^{2}(1-\frac{4m^{2}}{s}),
     \; \; \;    \vec \beta^{\,'}_{c} \cdot \vec
\beta^{\,'}_{d}=-\lambda_{2}^{2}(1-\frac{4m^{2}}{s}),
\nonumber\\
&\ & \vec \beta^{\,'}_{a} \cdot \vec \beta^{\,'}_{c}=- \vec \beta^{\,'}_{a}
\cdot \vec \beta^{\,'}_{d}=
\lambda_{1}\lambda_{2}\gamma(1-\frac{4m^{2}}{s}+\frac{2t}{s}).
\label{eq:expdef}
\end{eqnarray}
Notice that terms directly proportional to $1/q_{0}$ have exactly
cancelled and disappeared upon squaring the current.  If we set $M$ = 0
in  Eq. (\ref{eq:correctedotj}), we recover Eq.(3.6) of
Ref.~\cite{haglin93}.

We have calculated the ratio
$R_{0}=-J^{2}_{\rm virtual}\,/-J^{2}_{\rm real}$ with the
angular-averaged currents of Eq.
(\ref{eq:jmu1}) and Eq. (\ref{eq:jmu}).
This ratio depends on the invariant energy $s$, the four-momentum
transfer $t$, and the dilepton invariant mass and energy: $M$ and
$q_{0}$. For the
purpose of comparison, we
arbitrarily input some reasonable values and plot $R_0$ against the
virtual photon
three-momentum in Fig.
\ref{fig2}.  We find sizeable
deviations from 1 in this quantity.  The ratio for $p n$ processes is
sensitive to the value of $\theta_{\rm c.m.}$, the scattering angle in the
nucleon-nucleon centre of mass,  whereas it is
quite insensitive for the $p p$ case. Interference effects evidently play
a role here.
To better understand the effect
of the ``complete'' current (also labelled ``virtual'', above) on the
differential
cross section for dilepton
production, we integrate over $t$ and plot $R_{1}=(\frac{d\sigma}{dM^{2}
d^{3}q})_{\rm virtual}/(\frac{d\sigma}{dM^{2}d^{3}q})_{\rm real}$ as a
function of the c.m. virtual photon momentum $|\vec{q}\,|$ in Fig.
\ref{fig3}, for
different dilepton invariant masses.  We are now essentially comparing
the effect of
using Eq. (\ref{eq:jmu1}) over Eq. ({\ref{eq:jmu}) in a dilepton
production calculation. Also, to actually
compare two sets of calculations that have been done previously, the
numerator is derived from Eq. (\ref{eq:d6}), while the denominator is
derived from R\"uckl's approach.

Before discussing the results in Fig. \ref{fig3}, let us restate that
the net lepton
pair spectrum will depend on the details of the
differential nucleon-nucleon elastic cross-section, $d \sigma / d t$.
This is evident
from Eq. (\ref{eq:diff}) and Eq. (\ref{eq:d6}).  It thus follows
that the present analysis will have to be redone for different systems
at different
energies, with possibly different conclusions. For example,
quark-quark and pion-pion
virtual bremsstrahlung calculations
have attracted
some recent attention in connection with ultrarelativistic heavy-ion
collisions
\cite{haglin93,cley93}. The consequences of the present formalism on
such studies is
presently being analyzed \cite{rahma}.
It is thus imperative to have an accurate
parametrization for the elastic nucleon-nucleon differential cross
section $d\sigma/dt$ over the relevant range in $t$.
For $p n$ elastic collisions, $d\sigma/dt$ is nearly symmetric for
kinetic energy less than 1 GeV. But at higher energies, the observed
distributions are not symmetric about $\theta_{c.m.}=90^{\circ}$ but rather
develop a stronger forward peak. This asymmetry increases with the
scattering energy and can suppress the $p n$ bremsstrahlung contribution
to dilepton production by a factor of 4 at 4.9 GeV~\cite{law93,haglin94}.
A fairly detailed parametrization at 4.9 GeV kinetic
energy was used in Ref.~\cite{haglin94} to calculate the absolute
sizes of the  $p n$ and $p p$ bremsstrahlung contributions. For practical
reasons, we instead
adopt the parametrizations of Refs. \cite{law93} and \cite{rlh80} for the
$p n$ and $p p$
elastic differential cross sections.  These two functional forms can
fit the experimental data for energies up to 6 GeV with the necessary accuracy.

Analyzing Fig. \ref{fig3}, the ratio
$R_{1}$ remains 2/3 for small invariant
mass $M \leq$  10 MeV.  For higher invariant masses, $R_1$ is greater
than 2/3  for almost
all values of the virtual
photon momentum, in the case of $p n$ scattering.  Increasing $M$ in the
$p p$ case,
$R_{1}$ remains
still considerably smaller than 2/3 at low momenta and drastically
increases as
the three-momentum grows.  The low-momentum value of $R_1$ is thus
quite different in
the two reactions at hand.
It is not simple to draw clear physical conclusions for these behaviours,
but the main
point is that these features
suggest that using the ``complete'' current for virtual photons
changes the
distribution of the dielectrons in phase space, especially in
the $p p$ case.

Upon integration, we arrive at
differential cross sections in terms of lepton pair invariant mass.
In Fig. \ref{fig4},
we show two curves:
the dashed curve corresponds to using the R\"uckl formula with the real
photon current,
Eq. (\ref{eq:jmu}). The solid curve corresponds to using Eq.
(\ref{eq:d6}) with the
current of Eq. (\ref{eq:jmu1}).
In each panel
three kinetic energies are shown:  1.0, 3.0 and 4.9 GeV. We are thus in effect
evaluating the difference between two approaches to calculating the
dielectron production in
nucleon-nucleon collisions: the one based on R\"uckl's equation and the
real photon
current (we remind the reader that this combination has become well known
and widely
used), and the one based on the more recent Lorentz-covariant formalism
with a more
general electromagnetic current for hadrons.
At these energies (which correspond to DLS measurements) one realizes
that the Lorentz-covariant  and
gauge-invariant  formul{\ae} that we adopted in this work
(solid curves) reduce
the dilepton production cross section only slightly for $M <0.5$ GeV in both
$p n$ and $p p$  processes compared with previous results.
For $M > 0.5$ GeV, the situation is a little
more complicated as a slight decrease or a slight increase is observed,
depending on the
beam energy.  We have verified that the ratio
$(\frac{d\sigma}{dM^{2}})_{\rm virtual}/
(\frac{d\sigma}{dM^{2}})_{\rm real}$ is equal to 2/3
in the small $M$ region for both $p n$ and $p p$ cases,
and found it to increases monotonically with $M$ for $p n$ process,
but decreases
monotonically for $p p$ as one goes toward the kinematical limit.
Recall that other physical processes
dominate the  dielectron yield for $M > 0.5$ GeV at  large
enough beam energy \cite{haglin94}.  In the large invariant mass region,
Fig. 4
suggests that
the previous bremsstrahlung calculations underestimate the dilepton yields
at most by a factor of 2 even at the kinetic energy 4.9 GeV, even though
our formalism
is not strictly applicable there.   For $ p p $
scattering, it is worthwhile to point out that the radiation intensity
is strongly depleted due to interference effects near the maximum $M$.

Thus our
findings at this point are essentially these:
the ``improved'' formalism does not affect
very much  the soft lepton
yields previously calculated. We do not address in detail the issue of
hard leptons
($M \geq 0.5 $ GeV) because
there, the present formalism is inadequate and also because the signal
is dominated by
other sources.
In the small invariant
mass region $M<0.5$ GeV, we find that this new formalism only  amounts to
the numerical difference of 2/3.

We have made above a formal comparison of two approaches. However, we
did discover that
the two formalisms would produce different phase space distributions of
lepton pairs:
see Fig. \ref{fig3}. Let us now consider the specific case of the DLS data.
It is a known fact that any theoretical calculation attempting to
reproduce the DLS data should first be filtered by the DLS experimental
acceptance.
It is then conceivable that the predictions of the two approaches could be
affected
quite differently by the experimental filter. We have run our two sets of
calculations
through the DLS experimental filter \cite{filter}.
In Fig. \ref{fig5} we
show the invariant mass dependence of the absolute differential
cross sections, at kinetic energies 1.0, 3.0 and 4.9 GeV.
This figure now
suggests that
the previous calculations overestimated the
low-mass dielectron yields by a factor of 2-3 for $n p$ and a factor of 2-5
for $p p$ bremsstrahlung in collisions at 4.9 GeV. Thus the effect of the DLS
acceptance is to make the differences between the two approaches more
pronounced. The
effect is smaller for lower
kinetic energies.
Definite conclusions for comparison with the data are presently being drawn as
many-body bremsstrahlung has been shown to be quantitatively important at
4.9 GeV \cite{haglin94}. Even though we feel that a complete reassessment
of previous
results is probably not warranted, surprises can not be excluded.

Another useful comparison is the relative radiative intensities of
$p p$ and $n p$
scattering. As mentioned above, $p p$ bremsstrahlung has often been
neglected because
of a classical multipole argument.
We calculate the ratio here, with the formalism based on the leading-term
approximation, {\em i.e.} with Eq. (\ref{eq:d6}) and Eq.
(\ref{eq:jmu1}). The relative
intensities depend on kinetic energy as shown in Fig. \ref{fig6} which presents
the ratio $R=(\frac{d\sigma}{dM^{2}})_{\rm pp}/(\frac{d\sigma}{dM^{2}})_{\rm
pn}$ as a function of invariant mass at different kinetic energies.
Proton-proton bremsstrahlung becomes more and more important as the kinetic
energy increases. At 2 GeV, for instance,  $R$ is nearly 0.5 for
small masses whereas at 4.9 GeV it becomes larger than 1.  We shall
return to this ratio in the next section.

\section{One step further: the exact treatment of phase space}
\label{phasesp}

In all of the above comparisons, we have consistently set $q = 0$ in the
4-dimensional phase space delta function. We now avoid making this
approximation and investigate the consequences. The many-body
Lorentz-invariant phase space can be expressed in terms of Mandelstam-type
invariants \cite{eb73}, or in this case one can perform the integrals
directly \cite{tit94}. Going back one step,
let's write the general equation that leads to Eq. (\ref{eq:diff}),
without its phase space correction. In the process $a + b \rightarrow c
+ d + e^{+}e^{-}$, the differential cross section for
lepton pair production with invariant mass $M$ and energy $q_0$ is
\begin{eqnarray}
E_{+}E_{-}\frac{d^{6}\sigma^{e^{+}e^{-}}_{ab\rightarrow
cd}}{d^{3}p_{+}d^{3}p_{-
}}
&=&
\frac{1}{4E_{a}E_{b}|{\bf v}_{a}-{\bf
v}_{b}|}\frac{\alpha^{2}}{8\pi^{4}} \frac{1}{M^{2}} \int
[-J^{2}-\frac{1}{M^{2}}(l \cdot J)^{2}]
 \nonumber\\
&\ &
\times\
\overline{|{\cal M}_{0}|^{2}}
(2\pi)^{4}\delta^{4}\left(p_{a}+p_{b}-p_{c}-p_{d}-q\right)
\frac{d^{3}p_{c}}
{(2\pi)^{3}2E_{c}}\frac{d^{3}p_{d}}{(2\pi)^{3}2E_{d}}\ ,
\label{eq:diff11}
\end{eqnarray}
where $\overline{|{\cal{M}}_{0}|^{2}}$ is the on-shell matrix element for the
scattering $ a + b \rightarrow c + d$, squared, summed over final spins
and averaged over initial spins. The differential cross section can be
rewritten as
\begin{eqnarray}
q_{0}\frac{d^{4}\sigma^{e^{+}e^{-}}_{ab \rightarrow cd}}{dM^{2}d^{3}q}
&=&
\frac{1}{4E_{a}E_{b}|{\bf v}_{a}-{\bf
v}_{b}|}\frac{\alpha^{2}}{48\pi^{5}}
\frac{1}{M^{2}}(1+\frac{2\mu^{2}}{M^{2}}) \sqrt{1-
\frac{4\mu^{2}}{M^{2}}} \nonumber\\
&\ &
\times \ \int (-J^{2})
\overline{|{\cal M}_{0}|^{2}} \delta^{4}\left(p_{a}+p_{b}-p_{c}-p_{d}-q\right)
\frac{d^{3}p_{c}}
{2E_{c}}\frac{d^{3}p_{d}}{2E_{d}}\ ,
\label{eq:diff12}
\end{eqnarray}
where $\mu$ is the rest mass of an individual lepton. In the $a + b$
centre of mass frame, one may perform some of the
integrals and get
\begin{eqnarray}
\int\delta^{4}\left(p_{a}+p_{b}-p_{c}-p_{d}-q\right)\frac{d^{3}p_{c}}
{2E_{c}}\frac{d^{3}p_{d}}{2E_{d}} &=&
\frac{1}{4}\int
\delta\left(\cos\theta_{p_{d}q}-\frac{s+M^{2}+2E_{d}q_{0}-2\sqrt{s}
(E_{d}+q_{0})}{2|\vec{p_{d}}||\vec{q}|}\right)  \nonumber\\
&\ &
\times\ \frac{1}{|\vec{q}|}dE_{d}d\cos\theta_{p_{d}q}d\phi_{p_d} \ .
\label{eq:diff13}
\end{eqnarray}
Furthermore, in the on-shell limit,  the squared matrix element summed
over final spins and averaged
over initial spins can be
related to the on-shell differential elastic cross section by
\begin{equation}
\overline{|{\cal M}_{0}|^{2}} = 16 \pi s(s-4m^{2})\frac{d\sigma}{dt}\ .
\end{equation}
Integrating further, one obtains
\begin{eqnarray}
\frac{d\sigma^{e^{+}e^{-}}}{dM^{2}} &=&
\frac{\alpha^{2}}{24\pi^{4}}
\frac{1}{M^{2}}(1+\frac{2\mu^{2}}{M^{2}}) \sqrt{1-
\frac{4\mu^{2}}{M^{2}}} \sqrt{s(s-4m^{2})} \int
(-J^{2})\frac{d\sigma_{ab\rightarrow
cd}}{dt} \nonumber\\
&\ &
\times\ \delta\left(\cos\theta_{p_{d}q}-\frac{s+M^{2}+2E_{d}q_{0}-2\sqrt{s}
(E_{d}+q_{0})}{2|\vec{p_{d}}||\vec{q}|}\right) \nonumber \\
&\ &
dq_{0} dE_{d} d\Omega_{q}
d\cos\theta_{p_{d}q} d\phi_{p_d} \ .
\label{eq:diff14}
\end{eqnarray}
The range of the integration variables is such that the condition
\begin{equation}
|\cos\theta_{p_{d}q}|=\left|\frac{s+M^{2}+2E_{d}q_{0}-2\sqrt{s}
(E_{d}+q_{0})}{2|\vec{p_{d}}||\vec{q}|}\right| \leq 1
\end{equation}
has to be satisfied.

With the help of gauge invariance, it is not very difficult to write the
4-current squared in terms of ten scalar products $p_{i}\cdot p_{j}$.
Setting $m_{a}=m_{b}=m_{c}=m_{d}=m $, we obtain
\begin{eqnarray}
-J^{2} &=& -\frac{Q_{a}^{2}(4m^{2}-M^{2})}{(2p_{a}\cdot q-M^{2})^{2}}
           -\frac{Q_{b}^{2}(4m^{2}-M^{2})}{(2p_{b}\cdot q-M^{2})^{2}}
           -\frac{Q_{c}^{2}(4m^{2}-M^{2})}{(2p_{c}\cdot q+M^{2})^{2}}
           -\frac{Q_{d}^{2}(4m^{2}-M^{2})}{(2p_{d}\cdot q+M^{2})^{2}}
\nonumber\\
&\ &
-\frac{2Q_{a}Q_{b}(4p_{a}\cdot p_{b}-M^{2})}{(2p_{a}\cdot q
-M^{2})(2p_{b}\cdot
q
-M^{2})}
+\frac{2Q_{a}Q_{c}(4p_{a}\cdot p_{c}+M^{2})}{(2p_{a}\cdot q
-M^{2})(2p_{c}\cdot
q
+M^{2})}  \nonumber\\
&\ &
+\frac{2Q_{a}Q_{d}(4p_{a}\cdot p_{d}+M^{2})}{(2p_{a}\cdot q
-M^{2})(2p_{d}\cdot
q
+M^{2})}
+\frac{2Q_{b}Q_{c}(4p_{b}\cdot p_{c}+M^{2})}{(2p_{b}\cdot q
-M^{2})(2p_{c}\cdot
q
+M^{2})}
\nonumber\\
&\ &
+\frac{2Q_{b}Q_{d}(4p_{b}\cdot p_{d}+M^{2})}{(2p_{b}\cdot q
-M^{2})(2p_{d}\cdot
q
+M^{2})}
-\frac{2Q_{c}Q_{d}(4p_{c}\cdot p_{d}-M^{2})}{(2p_{c}\cdot q
+M^{2})(2p_{d}\cdot
q
+M^{2})} \ .
\label{eq:diff15}
\end{eqnarray}
In the $a + b$ centre of mass, the scalar products can be written in
terms of the integration variables of Eq. (\ref{eq:diff14}). For instance,
writing  $|\vec{p_{a}}|=|\vec{p_{b}} | = | \vec{p} | = \sqrt{s/4-m^{2}}$
and the energy  $E_{a}=E_{b}=E=\sqrt{s}/2$, then $p_{a}\cdot
p_{b}=E^{2}+ |\vec{p}|^{2};\ p_{a}\cdot q
=Eq_{0}-|\vec{p}| |\vec{q}|\cos\theta_{p_{a}q};\ p_{d}\cdot q
=E_{d}q_{0}-|\vec{p_{d}}|\vec{q}|\cos\theta_{p_{d}q}; $ and $p_{a}\cdot
p_{d}=EE_{d}-|\vec{p}| |\vec{p_{d}}|(\cos\theta_{p_{a}q}\cos\theta_{p_{d}q}
+\sin\theta_{p_{a }q} \sin\theta_{p_{d}q}\sin\phi_{p_d})$. All other
scalar products involve the above four.

On Fig. \ref{fig7} we display three different curves for the cases
of $p n$ and $p p $ scattering, at energies of 1.0, 3.0 and 4.9 GeV. We
compare calculations done with Eq. (\ref{eq:diff14}), calculations
done using R\"uckl's formula
with the current of Eq. (\ref{eq:jmu}), and calculations done using Eq.
(\ref{eq:d6}) together with the current of Eq.
(\ref{eq:jmu1}).
The three approaches display behaviors that are quite similar and are
not really remarkably different. The largest deviations occur
 at the lowest bombarding energies at low $M$. Recall that the ``soft
photon limit'' is not really properly defined by $M \rightarrow 0$.
Running our results through the DLS acceptance filter, in Fig. \ref{fig8},
one realizes that the results obtained with Eq. (\ref{eq:diff14}), and
those obtained with Eq. (\ref{eq:d6}) together with the current of Eq.
(\ref{eq:jmu1}) are very similar for the $p n$ reactions. The only
exception there  is the low energy
$p n$ spectrum at low invariant masses. In the case of $p p$, the filter
effects are such that the results with the complete phase space are lower than
the previous ones at high invariant masses.  The effect of the DLS
acceptance brings us to the conclusion that the position of the lepton
pairs in phase space is mostly determined by the nature of the
electromagnetic current, once the on-shell nucleon-nucleon cross
section is fixed. The $p p/n p$ ratio does however show some sensitivity
to the treatment of phase space, as displayed in Fig. \ref{fig7b}.

 There is no question that the current of Eq.
(\ref{eq:jmu1}) is more general and appropriate for leptons than that of Eq.
(\ref{eq:jmu}). In a given reaction the differences in dilepton
invariant cross sections,
before filtering,  that follow the
use of those two currents are not minuscule but can   not be called
spectacular. There is an observable that has a remarkably
different  behavior, whether or not the current (\ref{eq:jmu1}) is used in
calculations. We elaborate more on this in the following section.

\section{The angular distribution of bremsstrahlung dileptons}
\label{angular}
We have already mentioned that the nature of the electromagnetic current
for the bremsstrahlung producing hadrons will greatly affect the
position of the leptons in phase space. This already became clear in our
analysis of Fig. \ref{fig3}. Thus, it is very reasonable that the
differences in currents could be highlighted in a treatment focusing on
the fine points of angular distributions, for example. Angular
anisotropies have recently been put forward as a means of distinguishing
between competing lepton pair production sources \cite{brat94}. This
argument has power only if the angular distributions of those sources
can reliably be calculated. Let's elaborate on this below.

Owing to collision dynamics, the polarization of the virtual photon
eventually converting into a lepton pair may be such that, in the rest
frame of the dilepton, the single lepton distribution may not be
isotropic. This is the essence of the idea. Following \cite{brat94}, we
write the differential cross section for emission of a lepton pair of
invariant mass $M$, with a lepton coming out at a polar angle $\theta$
in the rest frame of the lepton pair as
\begin{equation}
S ( M, \theta) = {{d \sigma}\over {d M^2 d \cos \theta}} = A ( 1 + B
\cos^2 \theta)\ .
\label{eq:S}
\end{equation}
This enables us to write the polar anisotropy coefficient, $B$, as
\begin{equation}
B = {{S (M, \theta = 0^\circ )}\over{S (M, \theta = 90^\circ )}} - 1\ .
\end{equation}

Since the full phase is obviously crucial to the proper kinematics,
we use the
methods of the previous section. The cross section to be used in the
definition of $S$, above, is obtained by noting that for
electron-positron pair production \cite{pl94},
\begin{equation}
q_0 {{d^6 \sigma^{e^+ e^-}}\over{d M^2 d^3 q d \tilde{\Omega}_+}} =
{{1}\over{4}} E_+ E_- {{d^6 \sigma^{e^+ e^-}}\over{d^3 p_+ d^3 p_-}}\ ,
\end{equation}
where $q_0$ is the lepton pair energy and $\tilde{\Omega}_+$ is the
solid angle element for positron emission in the rest frame of the
lepton pair. With the help of the above equation, one can numerically
integrate Eq. (\ref{eq:diff11}) into the required format. In the
on-shell approximation we are using for the strong matrix element,
where ${\cal M}_0$ has no
dependence on $q$, the anisotropy coefficient $B$ only depends on
kinematics and on the electromagnetic current. Details of the strong
interaction do not influence it. We have numerically confirmed this by
varying the strong interaction differential cross section and observing
the constancy of $B$.

We first investigate the polar anisotropy of the lepton distribution
using the current associated with Eq. (\ref{eq:jmu}). The results are
shown in Fig. \ref{fig9}. We plot the coefficient $B$ as a function of
invariant mass, for different incident kinetic energies. Those energies
are 1, 2, 3, and 5 GeV. They can be readily identified in the figure by
their kinematical limit moving to higher invariant mass as the beam
energy grows. The calculations are for bremsstrahlung in $p n$
collisions. Our results at 1 and 2 GeV  reproduce those of Ref.
\cite{brat94}. Repeating the calculation with the full current of Eq.
(\ref{eq:jmu1}), we see our results change drastically, both
qualitatively and quantitatively. This is now plotted in Fig.
\ref{fig10}. The anisotropy coefficients no longer cross zero roughly
together at about $M = 0.25$ GeV. The crossing points have spread out,
and an important feature is that the curves have all been shifted upwards.
The anisotropy coefficients are now mostly positive, except in the high
invariant mass region of the last two incident kinetic energies. Also,
the minima have been shifted to higher invariant masses. Thus, it is
clear that any physical interpretation relying on the angular anisotropy
of the lepton spectrum will thus depend on the details of the
calculation. In this observable, the differences in electromagnetic
currents clearly stand out.

For the case of $p p$ bremsstrahlung, the use of the complete current is
absolutely crucial. The current of Eq. (\ref{eq:jmu}) is used to compute
the polar anisotropy coefficient plotted in Fig. \ref{fig11}. The
current of Eq. (\ref{eq:jmu1}) is used to compute the curves
appearing in Fig.
\ref{fig12}. The difference is striking. With the complete current, the
$p p$ signal bears some resemblance to its $p n$ counterpart, whereas the
current associated with real photons yields a complete different
picture, both qualitatively and quantitatively.

\section{Summary}

In summary, we have used a Lorentz-covariant and gauge-invariant formalism
to calculate lepton pair production cross sections via $p n$ and $p p$
bremsstrahlung,
in the leading term approximation.
The exact squared
current derived for virtual photons is suitable for any energy and for any
two-body equal-mass system.  With the appropriate generalizations one could
apply it to an unequal mass system and to reactions involving
many-body final states. This is important and should be done.
We find that using the complete current for virtual
photons will produce differential dilepton cross sections that are not
very different from those obtained
using the real
photon current.
Since the exact virtual photon current changes the momentum
distributions of the dileptons, the differences are somewhat accentuated
by running the
calculations through the DLS acceptance filter. However, the different phase
space population associated with the different currents emerge
in a much more striking fashion when plotting the polar
anisotropy coefficient for
leptons in the lepton pair rest frame. We feel that the lepton angular
anisotropy might in fact be useful in the identification of different
sources, but only once the calculations are deemed to possess the
required level of reliability and precision.

We do not suggest that
the DLS bremsstrahlung analyses have to be
completely redone. However, larger  differences in the
approaches (the one based on R\"uckl's formula and the Lorentz-covariant
one) do show up in the $p p$ channel, comparing with the $p n$
channel. Thus to pin down the various mechanisms for
dilepton production more precisely and to compare quantitatively to data, it
does seem necessary to scrutinize some of the previous
bremsstrahlung estimates,
as our results suggest that some of them might have somewhat
overestimated the low-mass yields.

The relative bremsstrahlung intensities from $p p$  and $p n$ scattering
have been
an actively debated issue for some time.  We provided a practical and
quantitative way to illustrate this, for kinetic energies up to 6 GeV.
Our results suggest that $p p$
bremsstrahlung is quite important when the kinetic energy is higher
than roughly 2 GeV.

\centerline{\bf Acknowledgments}
We are happy to acknowledge discussions and a useful correspondence
with P. Lichard. We thank him for suggesting changes in the first
version of this paper. We also acknowledge useful discussions
with H. Eggers and O. V. Teryaev.
This work has been supported in part by the Natural Sciences and
Engineering Research Council of Canada, by the FCAR fund of the Qu\'ebec
Government, by a NATO Collaborative Research Grant,  and by the National
Science
Foundation under grant number 94-03666.

\appendix
\section*{Dilepton radiation from four-fermion interactions}

In this appendix we derive the matrix element for lepton pair radiation
for a generic form of the strong four-fermion interaction, where we take
all the fermions to be charged for the sake of generality. We shall use
and generalize slightly the methods of Ref. \cite{pl94}. Our results are
in agreement with those of that reference.

We thus start with
4 diagrams  like that of Fig. \ref{fig1} where we consider
radiation from the external lines, in both the initial and final state.
Extra diagrams needed to antisymmetrize the final state can be added
without problems.  A
little more specifically, we consider a process
\begin{eqnarray}
a+b\rightarrow c+d+l^{+}+l^{-}
\end{eqnarray}
where {\it a,b,c,d} are all charged spin-1/2 fermions, and
$p_a + p_b = p_c + p_d
+ q$ with $q = p_+ + p_-$.  $\Gamma$ represents the
generic hadronic interaction
together with external line factors ($\bar{u}^{(s)}(p)$ or
$u^{(s)}(p)$) for one incoming and one outgoing fermion. For example,
$\Gamma_{ca}$ includes the factors $\bar{u}^{(s_{c})}(p_{c})$ and
$u^{(s_{a})}(p_{a})$, whereas $\Gamma_{db}$ includes the factors
$\bar{u}^{(s_{d})}(p_{d})$ and $u^{(s_{b})}(p_{b})$.
If we consider bremsstrahlung radiation from particle $a$, the
Feynman amplitude for that process can be written as
\begin{eqnarray}
{\cal M}_{a}
&=&
\bar{u}^{(s_{-})}(p_{-})(-i(-e)\gamma_{\nu})v^{(s_{+})}(p_{+})
\left(\frac{-ig^{\mu\nu}}{q^{2}}\right)
\\
&&
\ \times
\bar{u}^{(s_{c})}(p_{c})\Gamma_{db}(p_{a}-q)
\left(\frac{i}{\gamma\cdot(p_{a}-q)-m_{a}}\right)
\left(-iQ_{a}e\gamma_{\mu}\right)u^{(s_{a})}(p_{a}) \ .
\nonumber
\end{eqnarray}
Generally, $\Gamma$ depends on a four values of incoming and outgoing
four-momenta. We have made explicit its dependence on the four-momentum
involved with the emission of the virtual photon, only.
We can write
\begin{eqnarray}
{\cal M}_{a}
&=&
-Q_{a}eL_{\mu}
\bar{u}^{(s_{c})}(p_{c})\Gamma_{db}(p_{a}-q)
\left(
\frac{\gamma\cdot(p_{a}-q)+m_{a}}{2p_{a}\cdot q-M^{2}}
\right)
\gamma^{\mu}u^{(s_{a})}(p_{a})\ ,
\end{eqnarray}
where
\begin{eqnarray}
L_{\mu}
&=&
\frac{e}{M^{2}}{u}^{(s_{-})}(p_{-})\gamma_{\mu}v^{(s_{+})}(p_{+}) \ .
\end{eqnarray}
Similarly, for bremsstrahlung radiation from particle {\it b},{\it c}
and {\it d},
\begin{eqnarray}
{\cal M}_{b}
&=&
-Q_{b}eL_{\mu}
\bar{u}^{(s_{d})}(p_{d})\Gamma_{ca}(p_{b}-q)
\left(
\frac{\gamma\cdot(p_{b}-q)+m_{b}}{2p_{b}\cdot q-M^{2}}
\right)
\gamma^{\mu}u^{(s_{b})}(p_{b}) \ , \
\end{eqnarray}
\begin{eqnarray}
{\cal M}_{c}
&=&
Q_{c}eL_{\mu}
\bar{u}^{(s_{c})}(p_{c})\gamma^{\mu}
\left(
\frac{\gamma\cdot(p_{c}+q)+m_{c}}{2p_{c}\cdot q+M^{2}}
\right)
\Gamma_{db}(p_{c}+q)u^{(s_{a})}(p_{a}) \ , \
\end{eqnarray}
 and
\begin{eqnarray}
{\cal M}_{d}
&=&
Q_{d}eL_{\mu}
\bar{u}^{(s_{d})}(p_{d})\gamma^{\mu}
\left(
\frac{\gamma\cdot(p_{d}+q)+m_{d}}{2p_{d}\cdot q+M^{2}}
\right)
\Gamma_{ca}(p_{d}+q)u^{(s_{b})}(p_{b})\ .
\end{eqnarray}
Examine ${\cal M}_a$. Using the Dirac equation with gamma-matrix
algebra, one can show that
\begin{eqnarray}
\left(\gamma\cdot(p_{a}-q)+m_{a}\right)\gamma^{\mu}u^{(s_{a})}(p_{a})
&=&
\left(2p_{a}^{\mu}-q^{\mu}-\frac{1}{2}
\left[\gamma\cdot q,\gamma^{\mu}\right]\right)u^{(s_{a})}(p_{a}) \ .
\end{eqnarray}
Hence,
\begin{eqnarray}
{\cal M}_{a}
&=&
eL_{\mu}
\bar{u}^{(s_{c})}(p_{c})\Gamma_{db}(p_{a}-q)
\left(
-Q_{a}
\frac{\left(2p_{a}-q\right)^{\mu}}{2p_{a}\cdot q-M^{2}}
-
Q_{a}
\frac{\left[\gamma^{\mu},\gamma\cdot q\right]}
     {2\left(2p_{a}\cdot q-M^{2}\right)}
\right)
\nonumber\\
&&\ \times
u^{(s_{a})}(p_{a}) \ .
\end{eqnarray}
Similar expressions are derived for the other matrix elements. The
matrix element for reaction $a + b \rightarrow  c + d + l^+ + l^-$ is
\begin{eqnarray}
{\cal M}
&=&
{\cal M}_{a}+{\cal M}_{b}+{\cal M}_{c}+{\cal M}_{d} \ .
\end{eqnarray}
In order to identify the terms order-by-order in $q$, one needs to go to
next to leading order. This is because of the fact that in the case of
radiating fermions, the radiations from the external legs contain the
leading order part and also contain some next-to-leading order
contributions \cite{bpp91}. One then systematically has to go to the
next order to collect the missing pieces of the amplitude that will
restore gauge invariance.
To next-to-leading order one can then write
\begin{eqnarray}
\Gamma\left(p_{i}\pm q\right)
&=&
\Gamma\pm q^{\alpha}\frac{\partial\Gamma}{\partial p_{i}^{\alpha}}\ .
\end{eqnarray}
Also, in the spirit of the above discussion, we introduce the
electromagnetic contact term associated with the $i^{\rm th}$ fermion
line leaving the strong interaction core
\begin{eqnarray}
C_{i}^{\mu}&=&-eQ_{i}
\frac{\partial\Gamma}{\partial p_{i,\mu}} \ .
\end{eqnarray}
Then  we have, up to leading and next-to-leading order in {\it q},
\begin{eqnarray}
{\cal M}
&=&
J^{\mu}
\bar{u}^{(s_{c})}(p_{c})\Gamma_{db}u^{(s_{a})}(p_{a})
+
\bar{u}^{(s_{c})}(p_{c})K_{db}^{\mu}u^{(s_{a})}(p_{a})
\nonumber\\
&&
+\bar{u}^{(s_{d})}(p_{d})K_{ca}^{\mu}u^{(s_{b})}(p_{b})
\label{eq:appM}
\end{eqnarray}
with
\begin{eqnarray}
\bar{u}^{(s_{c})}(p_{c})\Gamma_{db}u^{(s_{a})}(p_{a})
\ =\
\bar{u}^{(s_{d})}(p_{d})\Gamma_{ca}u^{(s_{b})}(p_{b}) \ =\ {\cal M}_0 ( s_a,
s_b, s_c, s_d )\ .
\end{eqnarray}
In  Eq. (\ref{eq:appM}), the current $J^\mu$ is defined as in Eq.
(\ref{eq:jmu1}), and ${\cal M}_0 (s_a, s_b, s_c, s_d )$ is the matrix element
for the hadronic on-shell scattering where $p_a (s_a) + p_b (s_b)
\rightarrow p_c (s_c) + p_d (s_d)$. The momentum labelling of ${\cal M}_0$ has
been suppressed for simplicity.
Also,
\begin{eqnarray}
K_{db}^{\mu}
&=&
-\frac{Q_{a}}{4p_{a}\cdot q}
\Gamma_{db}\left[\gamma^{\mu},\gamma\cdot q\right]
+
\frac{Q_{c}}{4p_{c}\cdot q}
\left[\gamma^{\mu},\gamma\cdot q\right]\Gamma_{db}
\nonumber\\
&&
-Q_{a}\left(-g^{\mu\alpha}q^{\beta}+g^{\mu\beta}q^{\alpha}\right)
\frac{p_{a,\alpha}}{p_{a}\cdot q}
\frac{\partial\Gamma_{db}}{\partial p_{a}^{\beta}}
\nonumber\\
&&
+Q_{c}\left(g^{\mu\alpha}q^{\beta}-g^{\mu\beta}q^{\alpha}\right)
\frac{p_{c,\alpha}}{p_{c}\cdot q}
\frac{\partial\Gamma_{db}}{\partial p_{c}^{\beta}} \ ,
\end{eqnarray}
and
\begin{eqnarray}
K_{ca}^{\mu}
&=&
-\frac{Q_{b}}{4p_{b}\cdot q}
\Gamma_{ca}\left[\gamma^{\mu},\gamma\cdot q\right]
+
\frac{Q_{d}}{4p_{d}\cdot q}
\left[\gamma^{\mu},\gamma\cdot q\right]\Gamma_{ca}
\nonumber\\
&&
-Q_{b}\left(-g^{\mu\alpha}q^{\beta}+g^{\mu\beta}q^{\alpha}\right)
\frac{p_{b,\alpha}}{p_{b}\cdot q}
\frac{\partial\Gamma_{ca}}{\partial p_{b}^{\beta}}
\nonumber\\
&&
+Q_{d}\left(g^{\mu\alpha}q^{\beta}-g^{\mu\beta}q^{\alpha}\right)
\frac{p_{d,\alpha}}{p_{d}\cdot q}
\frac{\partial\Gamma_{ca}}{\partial p_{d}^{\beta}} \ .
\end{eqnarray}
Squaring the net matrix element,
we sum over the final state spins and average over the initial
state ones and obtain
\begin{eqnarray}
\frac{1}{4}\sum_{s_{+},s_{-},s_{a},s_{b},s_{c},s_{d}}
\left|{\cal M}\right|^{2}
&=&
e^2 \left(
J^{\mu}J^{\nu}\overline{\left|{\cal M}_{0}\right|^{2}}
+
H^{\mu\nu}
\right) \sum_{s_{+},s_{-}}L_{\mu}
L^{\dag}_{\nu}\ =\ \overline{\left| {\cal M} \right|^2}\ ,
\end{eqnarray}
where
\begin{eqnarray}
\frac{1}{4} \sum_{s_a, s_b, s_c, s_d} \left| {\cal M}_0 \right|^2 \ =\
\overline{\left| {\cal M}_0 \right|^2}\ ,
\end{eqnarray}
\begin{eqnarray}
H^{\mu\nu}
&=&
J^{\mu}
\left[
\frac{1}{2}\left(A^{\nu}+A^{\dagger\nu}\right)
+
\frac{1}{2}\left(B^{\nu}+B^{\dagger\nu}\right)
\right]
+
(\mu\longleftrightarrow\nu) \ ,
\end{eqnarray}
\begin{eqnarray}
A^{\nu}
\ =\
\frac{1}{4}
{\rm Tr}
\left[
\left(\gamma\cdot p_{c}+m_{c}\right)\Gamma_{db}
\left(\gamma\cdot p_{a}+m_{a}\right)K_{db}'^{\nu}
\right] \ ,
\
\end{eqnarray}
\begin{eqnarray}
A^{\dagger\nu}
\ =\
\frac{1}{4}
{\rm Tr}
\left[
\left(\gamma\cdot p_{c}+m_{c}\right)K_{db}^{\nu}
\left(\gamma\cdot p_{a}+m_{a}\right)\Gamma'_{db}
\right] \ ,
\
\end{eqnarray}
and
\begin{eqnarray}
B^{\nu}
\ =\
\frac{1}{4}
{\rm Tr}
\left[
\left(\gamma\cdot p_{d}+m_{d}\right)\Gamma_{ca}
\left(\gamma\cdot p_{b}+m_{b}\right)K_{ca}'^{\nu}
\right] \ .
\end{eqnarray}
We have further defined
\begin{eqnarray}
K_{db}'^{\mu}
&=&
-\frac{Q_{a}}{4p_{a}\cdot q}
\left[\gamma\cdot q,\gamma^{\mu}\right]
\Gamma'_{db}
+
\frac{Q_{c}}{4p_{c}\cdot q}\Gamma'_{db}
\left[\gamma\cdot q,\gamma^{\mu}\right]
\nonumber\\
&&
-Q_{a}\left(-g^{\mu\alpha}q^{\beta}+g^{\mu\beta}q^{\alpha}\right)
\frac{p_{a,\alpha}}{p_{a}\cdot q}
\frac{\partial\Gamma'_{db}}{\partial p_{a}^{\beta}}
\nonumber\\& &
+
Q_{c}\left(g^{\mu\alpha}q^{\beta}-g^{\mu\beta}q^{\alpha}\right)
\frac{p_{c,\alpha}}{p_{c}\cdot q}
\frac{\partial\Gamma'_{db}}{\partial p_{c}^{\beta}} \ ,
\end{eqnarray}
and
\begin{eqnarray}
K_{ca}'^{\mu}
&=&
-\frac{Q_{b}}{4p_{b}\cdot q}
\left[\gamma\cdot q,\gamma^{\mu}\right]
\Gamma'_{ca}
+
\frac{Q_{d}}{4p_{d}\cdot q}
\Gamma'_{ca}
\left[\gamma\cdot q,\gamma^{\mu}\right]
\nonumber\\
&&
-
Q_{b}\left(-g^{\mu\alpha}q^{\beta}+g^{\mu\beta}q^{\alpha}
\right)\frac{p_{b,\alpha}}{p_{b}\cdot q}
\frac{\partial\Gamma'_{ca}}{\partial p_{b}^{\beta}}
\nonumber\\
&&
+
Q_{d}\left(g^{\mu\alpha}q^{\beta}-g^{\mu\beta}q^{\alpha}
\right)\frac{p_{d,\alpha}}{p_{d}\cdot q}
\frac{\partial\Gamma'_{ca}}{\partial p_{d}^{\beta}}\ ,
\end{eqnarray}
where
\begin{eqnarray}
C'&=&\gamma^{0}C^{\dag}\gamma^{0} \ .
\end{eqnarray}

Next, combining the traces, one can write
\begin{eqnarray}
&& {\rm Tr}
\left[  \left(\gamma\cdot p_{c}+m_{c}\right)\Gamma_{db}
\left(\gamma\cdot p_{a}+m_{a}\right) K_{db}'^{\nu}  \right]
+
{\rm Tr}
\left[  \left(\gamma\cdot p_{c}+m_{c}\right) K_{db}^{\nu}
\left(\gamma\cdot p_{a}+m_{a}\right)\Gamma'_{db}  \right]
\nonumber\\
&&\qquad
=D+E\ ,
\end{eqnarray}
where
\begin{eqnarray}
\lefteqn{
D\ =\ -\frac{Q_{a}}{p_{a}\cdot q}p_{a,\alpha}
\left(-g^{\nu\alpha}q^{\beta}+g^{\nu\beta}q^{\alpha}\right)
\ \times
}
\nonumber\\
&& \qquad
{\rm Tr}
\left[
\left(\gamma\cdot p_{c}+m_{c}\right)
\frac{\partial\Gamma_{db}} {\partial p_{a}^{\beta}}
\left(\gamma\cdot p_{a}+m_{a}\right)\Gamma'_{db}
+
\left(\gamma\cdot p_{c}+m_{c}\right)\Gamma_{db}
\gamma_{\beta}\Gamma'_{db}
\right.
\nonumber\\
&& \qquad\qquad
+
\left.
\left(\gamma\cdot p_{c}+m_{c}\right)\Gamma_{db}
\left(\gamma\cdot p_{a}+m_{a}\right)
\frac{\partial\Gamma'_{db}}{\partial p_{a}^{\beta}}
\right] \ ,
\end{eqnarray}
and
\begin{eqnarray}
\lefteqn{
E\ =\ \frac{Q_{c}}{p_{c}\cdot q}p_{c,\alpha}
\left(g^{\nu\alpha}q^{\beta}-g^{\nu\beta}q^{\alpha}\right)
\ \times
}
\nonumber\\
&& \qquad
{\rm Tr}
\left[
\Gamma'_{db}\gamma_{\beta}\Gamma_{db}
\left(\gamma\cdot p_{a}+m_{a}\right)
+
\left(\gamma\cdot p_{c}+m_{c}\right)
\frac{\partial\Gamma_{db}}{\partial p_{c}^{\beta}}
\left(\gamma\cdot p_{a}+m_{a}\right)\Gamma'_{db}
\right.
\nonumber\\
&& \qquad\qquad
+
\left.
\left(\gamma\cdot p_{c}+m_{c}\right)\Gamma_{db}
\left(\gamma\cdot p_{a}+m_{a}\right)
\frac{\partial\Gamma'_{db}}{\partial p_{c}^{\beta}}
\right] \ .
\end{eqnarray}
Similarly,
\begin{eqnarray}
&& {\rm Tr}
\left[  \left(\gamma\cdot p_{d}+m_{d}\right)\Gamma_{ca}
\left(\gamma\cdot p_{b}+m_{b}\right) K_{ca}'^{\nu}  \right]
+
{\rm Tr}
\left[  \left(\gamma\cdot p_{d}+m_{d}\right) K_{ca}^{\nu}
\left(\gamma\cdot p_{b}+m_{b}\right)\Gamma'_{ca}  \right]
\nonumber\\
&& \qquad
=F+G\ ,
\end{eqnarray}
where
\begin{eqnarray}
\lefteqn{
F\ =\ -\frac{Q_{b}}{p_{b}\cdot q}p_{b,\alpha}
\left(-g^{\nu\alpha}q^{\beta}+g^{\nu\beta}q^{\alpha}\right)
\ \times
}
\nonumber\\
&& \qquad
{\rm Tr}
\left[
\left(\gamma\cdot p_{d}+m_{d}\right)
\frac{\partial\Gamma_{ca}} {\partial p_{b}^{\beta}}
\left(\gamma\cdot p_{b}+m_{b}\right)\Gamma'_{ca}
+
\left(\gamma\cdot p_{d}+m_{d}\right)\Gamma_{ca}
\gamma_{\beta}\Gamma'_{ca}
\right.
\nonumber\\
&& \qquad\qquad
+
\left.
\left(\gamma\cdot p_{d}+m_{d}\right)\Gamma_{ca}
\left(\gamma\cdot p_{b}+m_{b}\right)
\frac{\partial\Gamma'_{ca}}{\partial p_{b}^{\beta}}
\right] \ ,
\end{eqnarray}
and
\begin{eqnarray}
\lefteqn{
G\ =\ \frac{Q_{d}}{p_{d}\cdot q}p_{d,\alpha}
\left(g^{\nu\alpha}q^{\beta}-g^{\nu\beta}q^{\alpha}\right)
\ \times
}
\nonumber\\
&& \qquad
{\rm Tr}
\left[
\Gamma'_{ca}\gamma_{\beta}\Gamma_{ca}
\left(\gamma\cdot p_{b}+m_{b}\right)
+
\left(\gamma\cdot p_{d}+m_{d}\right)
\frac{\partial\Gamma_{ca}}{\partial p_{d}^{\beta}}
\left(\gamma\cdot p_{b}+m_{b}\right)\Gamma'_{ca}
\right.
\nonumber\\
&& \qquad\qquad
\left.
+
\left(\gamma\cdot p_{d}+m_{d}\right)\Gamma_{ca}
\left(\gamma\cdot p_{b}+m_{b}\right)
\frac{\partial\Gamma'_{ca}}{\partial p_{d}^{\beta}}
\right] \ .
\end{eqnarray}
Now,
\begin{eqnarray}
\frac{1}{4}\sum_{s_{a},s_{b},s_{c},s_{d}}
\left|{\cal M}_0\right|^{2}
=R \ {\rm or}\ S
\end{eqnarray}
where
\begin{eqnarray}
R\ =\
\frac{1}{4}
{\rm Tr}
\left[
\left(\gamma\cdot p_{d}+m_{d}\right)\Gamma_{ca}
\left(\gamma\cdot p_{b}+m_{b}\right)\Gamma'_{ca}
\right] \ ,
\
\end{eqnarray}
and
\begin{eqnarray}
S\ =\
\frac{1}{4}
{\rm Tr}
\left[
\left(\gamma\cdot p_{c}+m_{c}\right)\Gamma_{db}
\left(\gamma\cdot p_{a}+m_{a}\right)\Gamma'_{db}
\right] \ .
\
\end{eqnarray}
Therefore,
\begin{eqnarray}
\frac{\partial\overline{\left|{\cal M}_0\right|^{2}}}
{\partial p_{a}^{\alpha}}
&=&
\frac{1}{4}
{\rm Tr}
\left[
\left(\gamma\cdot p_{c}+m_{c}\right)
\frac{\partial\Gamma_{db}}{\partial p_{a}^{\alpha}}
\left(\gamma\cdot p_{a}+m_{a}\right)\Gamma'_{db}
+
\left(\gamma\cdot p_{c}+m_{c}\right)
\Gamma_{db}\gamma_{\alpha}\Gamma'_{db}
\right.
\nonumber\\
&& \qquad\
\left.
+
\left(\gamma\cdot p_{c}+m_{c}\right)\Gamma_{db}
\left(\gamma\cdot p_{a}+m_{a}\right)
\frac{\partial\Gamma'_{db}}{\partial p_{a}^{\alpha}}
\right] \ ,
\end{eqnarray}
\begin{eqnarray}
\frac{\partial\overline{\left|{\cal M}_0\right|^{2}}}
{\partial p_{b}^{\alpha}}
&=&
\frac{1}{4}
{\rm Tr}
\left[
\left(\gamma\cdot p_{d}+m_{d}\right)
\frac{\partial\Gamma_{ca}}{\partial p_{b}^{\alpha}}
\left(\gamma\cdot p_{b}+m_{b}\right)\Gamma'_{ca}
+
\left(\gamma\cdot p_{d}+m_{d}\right)
\Gamma_{ca}\gamma_{\alpha}\Gamma'_{ca}
\right.
\nonumber\\
&& \qquad\ \
\left.
+
\left(\gamma\cdot p_{d}+m_{d}\right)\Gamma_{ca}
\left(\gamma\cdot p_{b}+m_{b}\right)
\frac{\partial\Gamma'_{ca}}{\partial p_{b}^{\alpha}}
\right] \ ,
\end{eqnarray}
\begin{eqnarray}
\frac{\partial\overline{\left|{\cal M}_0\right|^{2}}}
{\partial p_{c}^{\alpha}}
&=&
\frac{1}{4}
{\rm Tr}
\left[
\Gamma'_{db}\gamma_{\alpha}\Gamma_{db}
\left(\gamma\cdot p_{a}+m_{a}\right)
+
\left(\gamma\cdot p_{c}+m_{c}\right)
\frac{\partial\Gamma_{db}}{\partial p_{c}^{\alpha}}
\left(\gamma\cdot p_{a}+m_{a}\right)\Gamma'_{db}
\right.
\nonumber\\
&& \qquad\ \
\left.
+
\left(\gamma\cdot p_{c}+m_{c}\right)\Gamma_{db}
\left(\gamma\cdot p_{a}+m_{a}\right)
\frac{\partial\Gamma'_{db}}{\partial p_{c}^{\alpha}}
\right] \ ,
\end{eqnarray}
and
\begin{eqnarray}
\frac{\partial\overline{\left|{\cal M}_0\right|^{2}}}
{\partial p_{d}^{\alpha}}
&=&
\frac{1}{4}
{\rm Tr}
\left[
\Gamma'_{ca}\gamma_{\alpha}\Gamma_{ca}
\left(\gamma\cdot p_{d}+m_{d}\right)
+
\left(\gamma\cdot p_{d}+m_{d}\right)
\frac{\partial\Gamma_{ca}}{\partial p_{d}^{\alpha}}
\left(\gamma\cdot p_{b}+m_{b}\right)\Gamma'_{ca}
\right.
\nonumber\\
&& \qquad\ \
\left.
+
\left(\gamma\cdot p_{d}+m_{d}\right)\Gamma_{ca}
\left(\gamma\cdot p_{b}+m_{b}\right)
\frac{\partial\Gamma'_{ca}}{\partial p_{d}^{\alpha}}
\right] \ .
\end{eqnarray}
Finally,
\begin{eqnarray}
\overline{\left|{\cal M}\right|^{2}}
&=&
e^{2}\sum_{s_{+},s_{-}}L_{\mu}L^{\dag}_{\nu}
\LA{\{}
J^{\mu}J^{\nu}\overline{\left|{\cal M}_{0}\right|^{2}}
+
\frac{1}{2}\sum_{i,j}
\frac{Q_{i}Q'_{j}}{(p_{i}\cdot q)(p_{j}\cdot q)}p_{i,\alpha}
\frac{\partial\overline{\left|{\cal M}_0\right|^{2}}}
{\partial p_{i}^{\beta}}
\nonumber\\
&& \qquad
\ \times
\left[p_{j}^{\mu}
\left(g^{\nu\alpha}q^{\beta}-g^{\nu\beta}q^{\alpha}\right)
+(\mu\longleftrightarrow\nu)\right]
\LA{\}}
\end{eqnarray}
where the sum in the second term runs over all fermions and
$Q'_{j}\ =\ -Q_{j}$ for the incoming fermions and
$Q'_{j}\ =\ Q_{j}$ for the outgoing fermions. The first term will lead
over the second one, in an expansion in powers of the virtual photon
four-momentum. The above equation,
together with
\begin{equation}
\sum_{s_+, s_-} L_\mu L^{\dag}_\nu\ =\ L_{\mu \nu}\ ,
\end{equation}
proves Eq. (\ref{eq:M2}).

\clearpage

\clearpage
\begin{figure}
\caption{The leading contributions for dilepton radiation in
the reaction $a+b \rightarrow c+d + e^+ e^-$.  \label{fig1}}
\end{figure}
\begin{figure}
\caption{The ratio $R\equiv -J^{2}_{\rm virtual}\,/-J^{2}_{\rm real}$
and its dependence
on the center-of-mass
dilepton momentum $q = |\vec q\,|$ and hadron scattering angle $\theta$ for
(a) $p n$, and (b) $p p$ processes at kinetic energy 4.9 GeV and
invariant mass $M$ = 0.2 GeV. \label{fig2}}
\end{figure}
\begin{figure}
\caption{The ratio of the differential cross section $R_{1}\equiv
(\frac{d\sigma}{dM^{2}d^{3}q})_{\rm virtual}\,
/(\frac{d\sigma}{dM^{2}d^{3}q})_{\rm real}$ as a function of
the center-of-mass dilepton momentum $q$ for (a) $n p$, and (b) $p p$
processes, at kinetic energy 4.9 GeV and for dilepton invariant masses
$M$= 0.01, 0.2, 0.6 and 0.8 GeV, respectively.\label{fig3}}
\end{figure}
\begin{figure}
\caption{The differential cross section for dilepton production from (a)
$p n$ and (b) $p p$ collisions
versus the dilepton invariant mass $M$.  The solid lines
are results using the virtual photon formalism described in the text, the
dotted lines
are the results obtained using R\"uckl's formula together with the
current for real
photons.
Results for kinetic energies $E_{\rm kin}$= 1.0, 3.0, and 4.9 GeV are
shown. For clarity of
presentation, the
results have been multiplied by a scaling factor.    The scaling factor
is 1 at  1 GeV, 10 at  3.0 GeV and 100 at  4.9 GeV.\label{fig4}}
\end{figure}
\begin{figure}
\caption{The same as in Fig. 4, but with the correction of the acceptance
filter. The solid symbols are with the formalism of  Eq.
(\protect\ref{eq:d6}) together with the current of Eq.
(\protect\ref{eq:jmu1}),
while the open symbols are the results of using R\"uckl's formula together
with the current for real photons.
\label{fig5}}
\end{figure}
\begin{figure}
\caption{The ratio of dilepton production cross sections in   $p p$
and $ n p$ reactions, $R =
(\frac{d\sigma}{dM})_{\rm pp}/(\frac{d\sigma}{dM})_{\rm pn}$,  as a
function of
invariant mass $M$ at kinetic energies 1.0,
2.0, 3.0 and 4.9 GeV, from bottom to top. \label{fig6}}
\end{figure}
\begin{figure}
\caption{We plot the differential cross section for production of lepton
pairs of invariant $M$ in the cases of $pn$ (left panel) and $pp$
collisions (right panel) at incident kinetic energies of 1, 3, and 4.9
GeV (bottom to top curves, respectively). The full curves represent
calculations done with Eq. (\protect\ref{eq:diff14}), the
short-dashed curves are
generated with the formula of R\"uckl, together with the current for
real photons, and the long-dashed curves are done using Eq.
(\protect\ref{eq:d6}) together with the current of Eq.
(\protect\ref{eq:jmu1}). The scaling factors are as in Fig.
\protect\ref{fig4}.            \label{fig7}}
\end{figure}
\begin{figure}
\caption{The same as Fig. \protect\ref{fig7}, but with the DLS acceptance
corrections. The open triangles are the results of using R\"uckl's
formula together
with the current for real photons. The open circles are associated with
Eq. (\protect\ref{eq:d6}) together with the current of Eq.
(\protect\ref{eq:jmu1}). The solid triangles are generated with Eq.
(\protect\ref{eq:diff14}).  \label{fig8} }
\end{figure}
\begin{figure}
\caption{Same caption as Fig. \protect\ref{fig6}, only this time the complete
phase space approach of section \protect\ref{phasesp} is used. \label{fig7b}}
\end{figure}
\begin{figure}
\caption{We plot the polar anisotropy coefficient, as defined in the
text, against lepton pair invariant mass for incident kinetic energies
of 1, 2, 3, and 5 GeV in $p n$ collisions. We use the current of Eq.
(\protect\ref{eq:jmu}).          \label{fig9}}
\end{figure}
\begin{figure}
\caption{Same caption as Fig. \protect\ref{fig9}. We use here the current of
Eq.
(\protect\ref{eq:jmu1}).          \label{fig10}}
\end{figure}
\begin{figure}
\caption{Same caption as Fig. \protect\ref{fig9}, but for $p p$ collisions.
  \label{fig11}}
\end{figure}
\begin{figure}
\caption{Same caption as Fig. \protect\ref{fig10}, but for
$p p$ collisions.           \label{fig12}}
\end{figure}


\begin{thebibliography}{24}
\bibitem{filter}We have used DLS filter version 2.0, provided by the
experimental
collaboration.
\end{thebibliography}
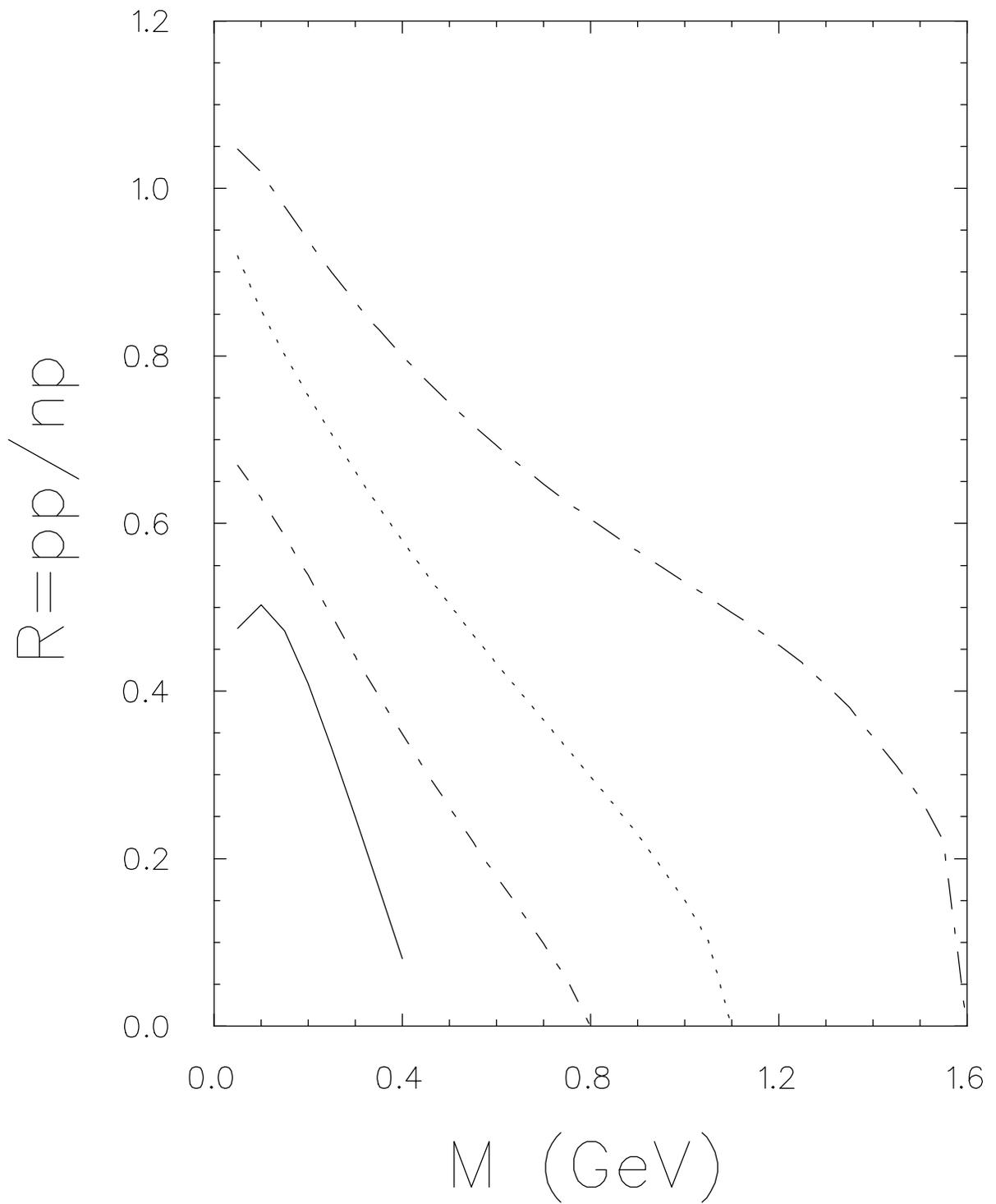
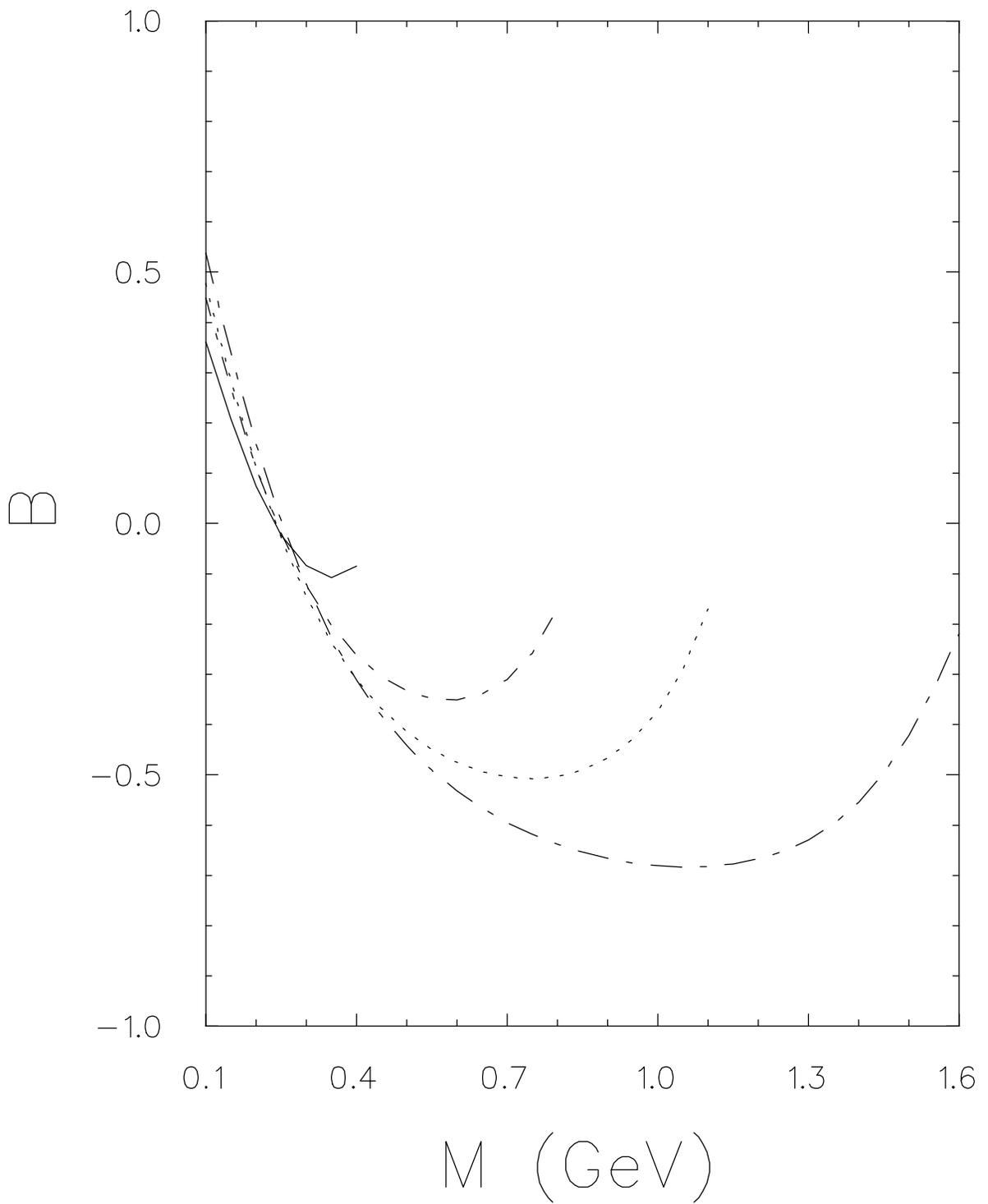
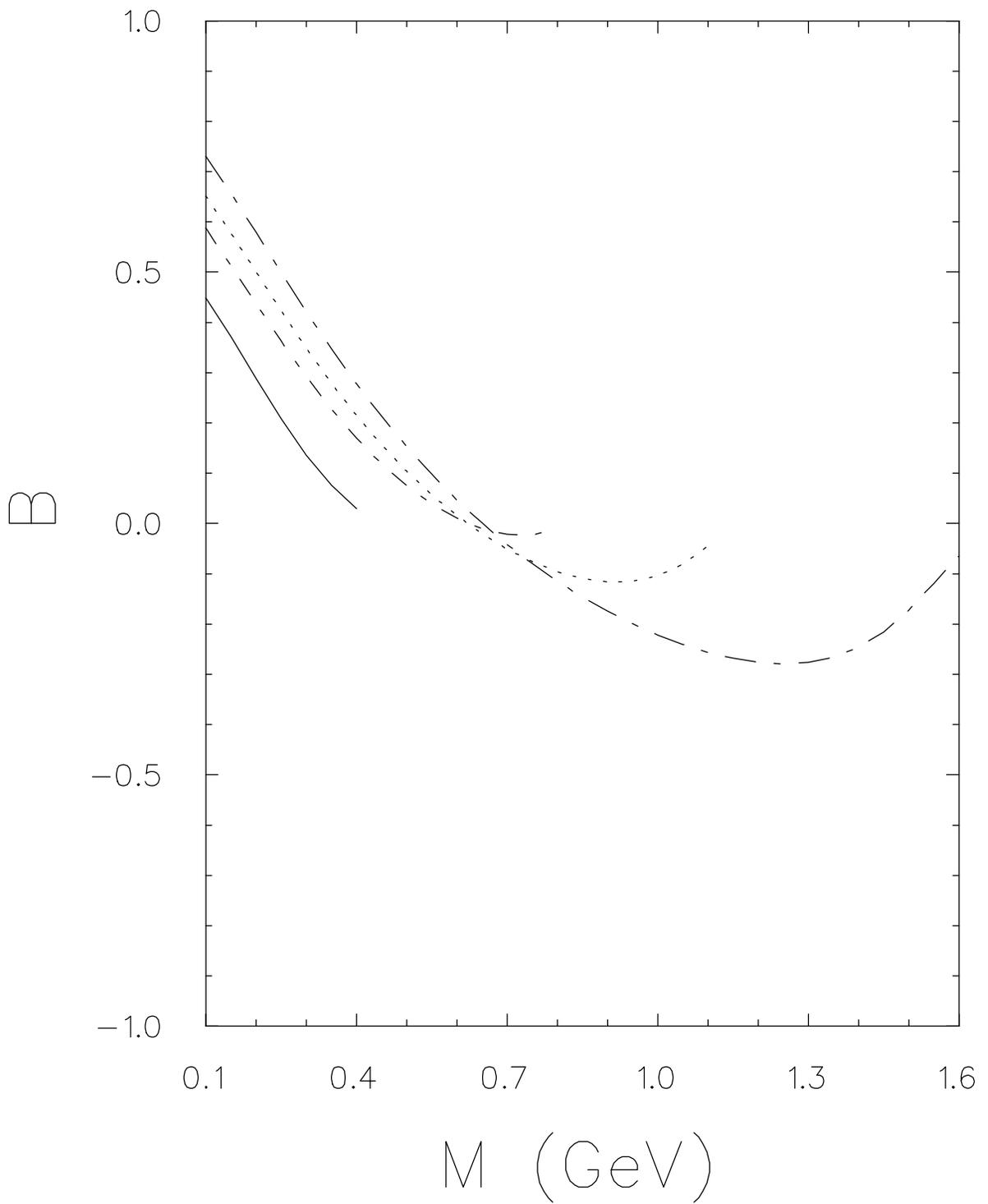
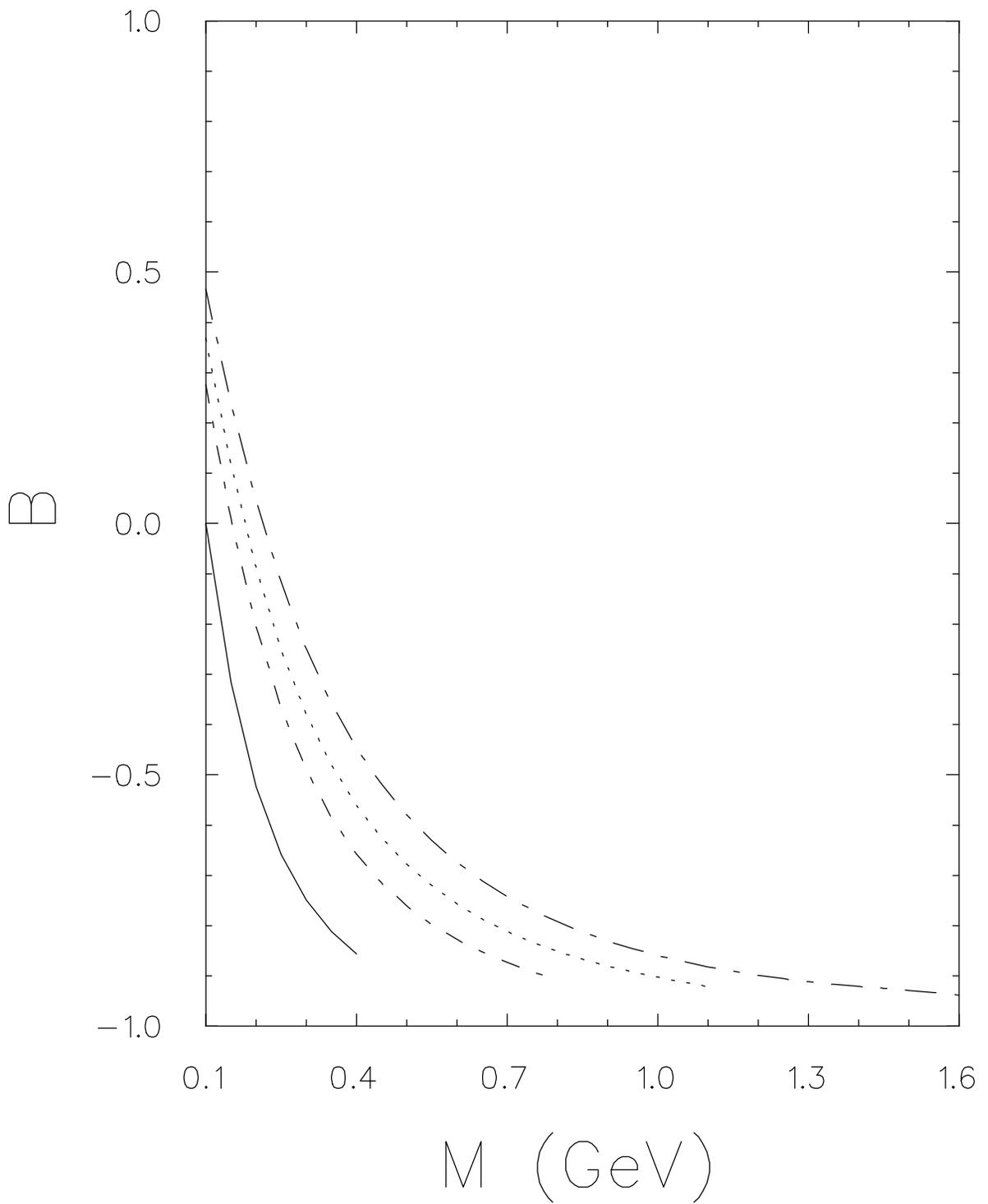
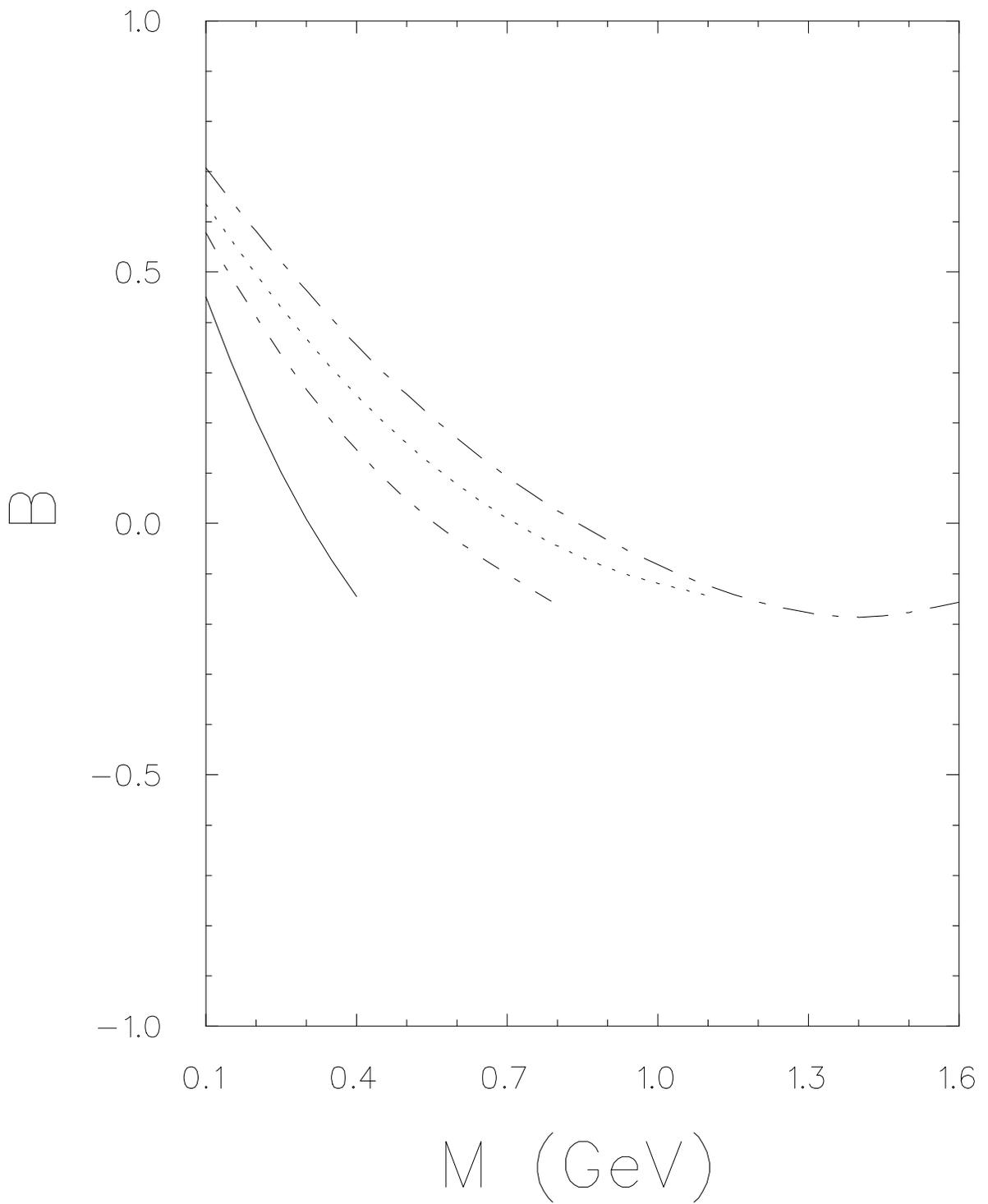
\end{document}